\documentclass[preprint,preprintnumbers,amsmath,amssymb]{revtex4}
\usepackage{graphicx}
\usepackage{color}
\usepackage{cancel}

\def\beq{\begin{equation}}
\def\eeq{\end{equation}}
\def\eeqn{\end{equation}}
\newcommand\iden{\leavevmode\hbox{\small1\normalsize\kern-.33em1}}

\newcommand{\bea} {\begin{eqnarray}}
\newcommand{\eea} {\end{eqnarray}}

\newcommand{\order}{{\cal O}}

\let\jnfont=\rm
\def\NPB#1,{{\jnfont Nucl.\ Phys.\ B }{\bf #1},}
\def\PLB#1,{{\jnfont Phys.\ Lett.\ B }{\bf #1},}
\def\EPJC#1,{{\jnfont Eur.\ Phys.\ Jour.\ C }{\bf #1},}
\def\PRD#1,{{\jnfont Phys.\ Rev.\ D }{\bf #1},}
\def\PRL#1,{{\jnfont Phys.\ Rev.\ Lett.\ }{\bf #1},}
\def\MPLA#1,{{\jnfont Mod.\ Phys.\ Lett.\ A }{\bf #1},}
\def\JPG#1,{{\jnfont J.\ Phys.\ G }{\bf #1},}
\def\CTP#1,{{\jnfont Commun.\ Theor.\ Phys.\ }{\bf #1},}
\def\JHEP#1,{{\jnfont JHEP \ }{\bf #1},}
\def\NPPS#1,{{\jnfont Nucl.\ Phys.\ Proc.\ Suppl.\ }{\bf #1},}
\def\CPC#1,{{\jnfont Computl.\ Phys.\ Commun.\ }{\bf #1},}
\def\CPL#1,{{\jnfont Chin.\ Phys.\ Lett. }{\bf #1},}
\def\AJS#1,{{\jnfont Astrophys.\ J.\ Suppl. }{\bf #1},}
\def\PR#1,{{\jnfont Phys.\ Rept. }{\bf #1},}
\def\AP#1,{{\jnfont Astropart.\ Phys. }{\bf #1},}
\def\EPL#1,{{\jnfont Europhys.\ Lett. }{\bf #1},}
\def\FP#1,{{\jnfont Fortsch.\ Phys. }{\bf #1},}
\def\JCAP#1 {{\jnfont JCAP \ }{\bf #1} }

\begin{document}

\title{95 GeV Higgs boson and nano-Hertz gravitational waves from domain walls in the N2HDM}
\renewcommand{\thefootnote}{\fnsymbol{footnote}}

\author{Haotian Xu, Yufei Wang, Xiao-Fang Han, Lei Wang$^{*}$\footnotetext{*) 
Corresponding author. Email address: leiwang@ytu.edu.cn}}
 \affiliation{$^1$Department of Physics, Yantai University, Yantai, Shandong 264005, China}
\renewcommand{\thefootnote}{\arabic{footnote}}

\begin{abstract}
We explore the diphoton and $b\bar{b}$ excesses at 95.4 GeV, as well as nano-Hertz gravitational waves originating from domain walls, 
within the framework of the next-to-two-Higgs-doublet model (N2HDM), which extends the two-Higgs-doublet model by introducing a real singlet scalar subject to a discrete 
$Z_2$ symmetry. The $Z_2$ symmetry is spontaneously broken by the non-zero vacuum expectation value of the singlet scalar, $v_s$, 
which leads to the formation of domain walls.
We discuss two different scenarios: in scenario A, the 95.4 GeV Higgs boson predominantly originates from the singlet field, 
while in scenario B, it arises mainly from the CP-even components of the Higgs doublets.
 Taking into account relevant theoretical and experimental constraints, 
we find that scenario A can fully account for both the diphoton and $b\bar{b}$ excesses at 95.4 GeV within the $1\sigma$ range. 
Moreover, the peak amplitude of the gravitational wave spectrum at a peak frequency of $10^{-9}$ Hz can reach $2 \times 10^{-12}$ for $v_s = 100$ TeV.
Scenario B only marginally accounts for
the diphoton and $b\bar{b}$ excesses at the $1\sigma$ level, but the peak amplitude of the gravitational wave spectrum at the peak frequency of $10^{-9}$ Hz
can reach $6\times 10^{-8}$ for $v_s=100$ TeV.
The nano-Hertz gravitational wave signals predicted in both scenarios can be tested by the
current and future pulsar timing array projects.

\end{abstract}
\maketitle

\section{Introduction} 
The possibility of additional Higgs states beyond the Standard Model (SM) is an open question.
The emergence of experimental anomalies hinting at a scalar resonance around 95.4 GeV
 has generated significant interest within the high-energy physics community.
Hints of a light scalar near 95.4 GeV first appeared in searches at LEP collider, where a 2.3$\sigma$ excess in the 
$e^+ e^-\to Z\phi$ process with $\phi\to b\bar{b}$ was reported \cite{LEPWorkingGroupforHiggsbosonsearches:2003ing}. 
Recently, the CMS collaboration has observed a local significance of 2.9$\sigma$
in the diphoton invariant mass spectrum around 95.4 GeV using full Run 2 data set of LHC \cite{cms95}. 
 The ATLAS experiment has also observed an excess in the diphoton channel, albeit with a lower significance \cite{atlas95}.
 Neglecting possible correlations, the
combined signal strength of ATLAS and CMS has a $3.1\sigma$ local excess in the diphoton channel around 95.4 GeV \cite{Biekotter:2023oen}.
There are numerous explanations on the excesses in the new physics models, for example
Refs. \cite{Biekotter:2023oen,Cao:2016uwt,Biekotter:2019kde,Heinemeyer:2021msz,Biekotter:2022jyr,
Aguilar-Saavedra:2023tql,Choi:2019yrv,Ma:2020mjz,Li:2022etb,Ellwanger:2023zjc,Dev:2023kzu,Bonilla:2023wok,Liu:2024cbr,Li:2023kbf,Abbas:2023bmm,
Ahriche:2023hho,Cao:2023gkc,Maniatis:2023aww,Belyaev:2023xnv,Azevedo:2023zkg,Bhatia:2022ugu,Chen:2023bqr,Ahriche:2023wkj,Arcadi:2023smv,
Borah:2023hqw,Banik:2023vxa,Dutta:2023cig,Bhattacharya:2023lmu,Ashanujjaman:2023etj,Escribano:2023hxj,Biekotter:2023jld,
Banik:2023ecr,Coloretti:2023wng,Ahriche:2022aoj,Benbrik:2022azi,Biekotter:2022abc,Biekotter:2021qbc,Abdelalim:2020xfk,Biekotter:2020cjs,
Aguilar-Saavedra:2020wrj,Cao:2019ofo,Kundu:2019nqo,Cao:2024axg,Kalinowski:2024uxe,Ellwanger:2024vvs,Benbrik:2024ptw,Lian:2024smg,
Yang:2024fol,Gao:2024qag,Khanna:2024bah,BrahimAit-Ouazghour:2024img,Gao:2024ljl,Mondal:2024obd,Banik:2024ugs,Hmissou:2025uep,Du:2025eop,Arhrib:2025pxy,Li:2025tkm,YaserAyazi:2024hpj}.

On the other hand, current observations from several pulsar timing array (PTA) collaborations have yielded strong statistical evidence 
for a stochastic gravitational wave background (SGWB) in the nano-Hertz regime, including NANOGrav \cite{NANOGrav:2023gor,NANOGrav:2023hfp}, 
the European Pulsar Timing Array (EPTA) \cite{EPTA:2023sfo,EPTA:2023akd}, 
the Parkes Pulsar Timing Array (PPTA) \cite{Reardon:2023gzh}, and the Chinese Pulsar Timing Array (CPTA) \cite{Xu:2023wog}. In addition to the supermassive black
hole binaries interpretation \cite{NANOGrav:2023hvm}, there are various cosmological sources of the gravitational waves (GWs), such as cosmic strings~\cite{Ellis:2023tsl,Bian:2023dnv,Wang:2023len,Lazarides:2023ksx,Eichhorn:2023gat,Chowdhury:2023opo,Antusch:2023zjk,Yamada:2023thl,Ge:2023rce,Basilakos:2023xof,Avgoustidis:2025svu}, domain walls~\cite{Kitajima:2023cek,Blasi:2023sej,Gouttenoire:2023ftk,Lu:2023mcz,Babichev:2023pbf,Gelmini:2023kvo,Guo:2023hyp,Zhang:2023nrs,Huang:2023zvs,Du:2023qvj,Li:2024psa,Li:2023gil,Lu:2024szr}, 
primordial first-order phase transitions~\cite{Fujikura:2023lkn,Franciolini:2023wjm,Bringmann:2023opz,Addazi:2023jvg,Bai:2023cqj,Han:2023olf,Zu:2023olm,Ghosh:2023aum,Xiao:2023dbb,
Li:2023bxy,DiBari:2023upq,Cruz:2023lnq,Gouttenoire:2023bqy,Ahmadvand:2023lpp,Salvio:2023ynn,Athron:2023mer,Jiang:2023qbm,He:2023ado,Chao:2023lox,Goncalves:2025uwh,Costa:2025csj,Salvio:2023blb},
 and inflation~\cite{Vagnozzi:2023lwo,Frosina:2023nxu,Liu:2023ymk,Unal:2023srk,Bari:2023rcw,Das:2023nmm,Jiang:2023gfe,Gorji:2023sil,An:2023jxf,Cang:2023ysz,Liu:2023pau,Liu:2023hpw}.

In this paper, we explore the possibility of simultaneously explaining the observed excess of 95.4 GeV Higgs boson and 
the nano-Hertz SGWB within the framework of the two-Higgs-doublet model extended by a real singlet scalar, commonly referred to as the N2HDM 
\cite{Biekotter:2019kde,Heinemeyer:2021msz,Biekotter:2022jyr,Chen:2013jvg,Muhlleitner:2016mzt,Engeln:2018mbg}. In this model, the 95.4 GeV Higgs boson 
arises from the mixing among three CP-even scalar fields, including the singlet field. 
In addition, the N2HDM respects a discrete $Z_2$ symmetry, which is spontaneously broken by the non-zero vacuum expectation value (VEV) of the singlet scalar.
The spontaneous breaking of this discrete $Z_2$ symmetry in the early Universe leads to the formation of domain walls that separate regions of distinct vacua.
The authors of \cite{Sassi:2024cyb} studied the possibility of restoring the electroweak symmetry in the N2HDM via the domain walls.
To ensure that the domain walls are unstable and eventually annihilate, the discrete symmetry is explicitly broken by introducing a very small bias term. 
The peak frequency of GWs produced by the domain walls is set by the time of annihilation, 
while the amplitude depends on the energy density of domain wall. This model has the potential to generate domain walls 
whose dynamics may produce GWs in the nano-Hertz frequency band, 
offering a possible explanation for the signals observed by PTA experiments.

The remainder of this paper is organized as follows. In Section II we present the key features of the N2HDM. 
In Sections III and IV, we examine the relevant theoretical and experimental constraints, and discuss the explanation of the diphoton and 
$b\bar{b}$ excesses at 95.4 GeV. In Section V we explore the domain walls and the associated GW spectra.  
 Finally, we summarize our conclusions in Section VI.

\section{The model} 
The two-Higgs-doublet model is extented by introducing a real singlet scalar, 
\bea
&&\Phi_1=\left(\begin{array}{c} \phi_1^+ \\
\frac{(v_1+\rho_1+i\eta_1)}{\sqrt{2}}\,
\end{array}\right)\,, 
\Phi_2=\left(\begin{array}{c} \phi_2^+ \\
\frac{(v_2+\rho_2+i\eta_2)}{\sqrt{2}}\,
\end{array}\right),
\Phi_S=\rho_s+v_s,
\eea
After spontaneous symmetry breaking, the fields $\Phi_1, \Phi_2$ and $\Phi_S$ acquire their VEVs
 $v_1$, $v_2$, and $v_s$, with $v\equiv \sqrt{v_1^2+v_2^2}=246$ GeV. The ratio of the two-Higgs-doublet VEVs is defined as $\tan\beta \equiv v_2 /v_1$.

The Higgs potential is written as
\begin{equation}
  \begin{aligned}
  V &= m_{11}^2 |\Phi_1|^2 + m_{22}^2 |\Phi_2|^2 - m_{12}^2 (\Phi_1^\dagger
  \Phi_2 + \text{h.c.}) + \frac{\lambda_1}{2} (\Phi_1^\dagger \Phi_1)^2 +
  \frac{\lambda_2}{2} (\Phi_2^\dagger \Phi_2)^2  \\
  &\hphantom{=} + \lambda_3
  (\Phi_1^\dagger \Phi_1) (\Phi_2^\dagger \Phi_2) + \lambda_4
  (\Phi_1^\dagger \Phi_2) (\Phi_2^\dagger \Phi_1) + \frac{\lambda_5}{2}
  [(\Phi_1^\dagger \Phi_2)^2 + \text{h.c.}]  \\
  &\hphantom{=} + \frac{m_S^2}{2}  \Phi_S^2 + \frac{\lambda_6}{8} \Phi_S^4 +
  \frac{\lambda_7}{2} (\Phi_1^\dagger \Phi_1) \Phi_S^2 +
  \frac{\lambda_8}{2} (\Phi_2^\dagger \Phi_2) \Phi_S^2.
\end{aligned}
  \label{eq:n2hdmpot}
\end{equation}
We assume that the model conserves CP-symmetry, meaning all coupling constants and mass parameters to be real. 
The Higgs potential in Eq. (\ref{eq:n2hdmpot}) respects a discrete $Z_2$ symmetry,
\begin{align}
  \Phi_1\to\Phi_1\,,\ \Phi_2\to\Phi_2\,,\ \Phi_S\to -\Phi_S\,, \label{eq:Z2}
\end{align}
which is spontaneously broken by the VEV of $\Phi_S$, giving rise to domain walls in the early Universe.
The conditions for minimizing the scalar potential lead to
\begin{align}
v_2 m_{12}^2 - v_1 m_{11}^2 &= \frac{1}{2} (v_1^3 \lambda_1 +
v_2^2 v_1 \lambda_{345} + v_s^2 v_1 \lambda_7)\,, \label{eq:n2hdmmin1} \\
v_1 m_{12}^2 - v_2 m_{22}^2 &= \frac{1}{2} (v_1^2 v_2 \lambda_{345} +
v_2^3 \lambda_2 + v_s^2 v_2 \lambda_8)\,, \label{eq:n2hdmmin2} \\
- v_s m_S^2 &= \frac{1}{2} v_s (v_1^2 \lambda_7 + v_2^2 \lambda_8 + v_s^2
\lambda_6)\,,  \label{eq:n2hdmmin3}
\end{align}
where
$\lambda_{345} \equiv \lambda_3 + \lambda_4 + \lambda_5$.

It is necessary to introduce an energy bias in the potential, which lifts the degenerate minima connected by the discrete $Z_2$ symmetry.
This bias destabilizes the domain walls, thereby evading the associated cosmological constraints \cite{Vilenkin:1981zs,Gelmini:1988sf,Larsson:1996sp}.
Here we introduce the $Z_2$ symmetry-breaking potential term by hand,
\beq
V_{\cancel{Z}_2}=-a_1 v_s^2 \Phi_S + \frac{a_1}{3} \Phi_S^3 \,.
\eeq
The total potential still has minima at $(v_1,v_2,\pm v_s)$, but there is an energy difference between them
\beq
V_{\rm bias}=V(v_1,v_2,-v_s)-V(v_1,v_2,v_s)=\frac{4}{3}a_1 v_s^3 \,.
\eeq
In fact, since the required values of $a_1$ is extremely small, we only consider it in
 the discussions of domain walls, and neglect it in other calculations.

Once spontaneous symmetry breaking occurs, the mass eigenstates are derived from the original fields through the application of rotation matrices
\begin{eqnarray}
\left(\begin{array}{c}h_1,h_2,h_3 \end{array}\right) =   \left(\begin{array}{c} \rho_1,  \rho_2,  \rho_s\end{array} \right) R^T, \\
\left(\begin{array}{c}G^0 \\ A \end{array}\right) =  \left(\begin{array}{cc}\cos\beta & \sin\beta \\ -\sin\beta & \cos\beta \end{array}\right)  \left(\begin{array}{c} \eta_1 \\ \eta_2 \end{array}\right) , \\
\left(\begin{array}{c}G^{\pm} \\ H^{\pm} \end{array}\right) =  \left(\begin{array}{cc}\cos\beta & \sin\beta \\ -\sin\beta & \cos\beta \end{array}\right)  \left(\begin{array}{c} \phi^{\pm}_1 \\ \phi^{\pm}_2 \end{array}\right),
\end{eqnarray}
with 
\beq
	R= \left( \begin{array}{*{20}{c}}
			c_{1}   c_{2} & s_{1}  c_{2}  & s_{2}\\
			  -  s_{1}  c_{3}-c_{1}   s_{2}   s_{3}  &  c_{1}  c_{3}-s_{1}  s_{2}  s_{3}   & c_{2}  s_{3}  \\
			 s_{ 1}  s_{3}-c_{1}  s_{2}  c_{3} &- s_{1}  s_{2}c_{3}  -c_{1}  s_{3}  & c_{2}  c_{3}  
	\end{array} \right).
\eeq
The shorthand notations are $s_{1,2,3}\equiv\sin\alpha_{1,2,3}$ and $c_{1,2,3}\equiv\cos\alpha_{1,2,3}$.
The $G^0$ and $G^\pm$ correspond to the Goldstone bosons, which are absorbed by the $Z$ and $W^\pm$. 
The remaining physical states consist of three CP-even scalars $h_{1,2,3}$,
 one pseudoscalars $A$, and a pair of charged scalar $H^{\pm}$.

The coupling constants in the Higgs potential can be expressed in terms of the scalar masses and mixing angles as
\begin{eqnarray}
&& \lambda_1  = \frac{ m_{h_1}^2 R_{11}^2+m_{h_2}^2 R_{21}^2 +m_{h_3}^2 R_{31}^2  - m_{12}^2 t_\beta}{v^2 c_\beta^2},  \nonumber \\  
&& \lambda_2 = \frac{  m_{h_1}^2 R_{12}^2 +m_{h_2}^2  R_{22}^2+m_{h_3}^2 R_{32}^2- m_{12}^2 t_\beta^{-1}}{v^2 s_\beta^2},  \nonumber \\  
&&\lambda_3 =  \frac{m_{h_1}^2 R_{11}R_{12}+m_{h_2}^2 R_{21}R_{22}+m_{h_3}^2 R_{31}R_{32}+ 2 m_{H^{\pm}}^2 s_\beta c_\beta - m_{12}^2}{v^2 s_\beta c_\beta },  \nonumber \\ 
&&\lambda_4 =  \frac{(m_A^2-2m_{H^{\pm}}^2 ) s_\beta c_\beta + m_{12}^2}{v^2 s_\beta c_\beta },   \nonumber \\    
&&\lambda_5=  \frac{ - m_A^2 s_\beta c_\beta  + m_{12}^2}{ v^2 s_\beta c_\beta },\nonumber \\ 
&&\lambda_6= \frac{m_{h_1}^2 R_{13}^2+m_{h_2}^2 R_{23}^2+m_{h_3}^2 R_{33}^2-2 a_1 v_s}{v_s^2} ,  \nonumber\\
&&\lambda_7= \frac{m_{h_1}^2 R_{11} R_{13}+m_{h_2}^2 R_{21} R_{23}+m_{h_3}^2 R_{31} R_{33}}{v v_s c_\beta},  \nonumber \\
&&\lambda_8=\frac{m_{h_1}^2 R_{12} R_{13}+m_{h_2}^2 R_{22} R_{23}+m_{h_3}^2 R_{32} R_{33}}{v v_s s_\beta},  \nonumber\\
\label{eq:lambdas}
\end{eqnarray}
with shorthand notation $t_{\beta}\equiv\tan\beta$.

In order to avoid the flavour changing neutral current at the tree-level, one imposes a $\mathbb{Z'}_2$ symmetry (see, e.g., the recent review \cite{Wang:2022yhm}), 
\beq
  u_R\to -u_R\,,\ d_R\to-d_R\,,\ e_R\to -e_R\,, \Phi_2\to -\Phi_2,\label{eq:Z2}
\eeq
while the other fields remain unchanged.
This results in the well-known type-I Yukawa interactions,
 \bea
- {\cal L} &=&Y_{u2}\,\overline{Q}_L \, \tilde{{ \Phi}}_2 \,u_R
+\,Y_{d2}\,
\overline{Q}_L\,{\Phi}_2 \, d_R\, + \, Y_{\ell 2}\,\overline{L}_L \, {\Phi}_2\,e_R+\, \mbox{h.c.}\,. \eea
Where $Q_L^T=(u_L\,,d_L)$, $L_L^T=(\nu_L\,,l_L)$,
$\widetilde\Phi_{2}=i\tau_2 \Phi_{2}^*$, and $Y_{u2}$,
$Y_{d2}$ and $Y_{\ell 2}$ are $3 \times 3$ matrices in family
space. However, the $\mathbb{Z'}_2$ symmetry is explicitly broken in the scalar potential of Eq. (\ref{eq:n2hdmpot}).

The neutral Higgs boson couplings, normalized to the SM values, are given by
\bea\label{hffcoupling} &&
y^{h_1}_V=c_2 c_{\beta 1},~y_{f}^{h_1}=c_2 \left(c_{\beta 1}- s_{\beta 1}\kappa_f\right), \nonumber\\
&&y^{h_2}_V = c_3 s_{\beta 1} - s_2 s_3 c_{\beta 1},~y^{h_2}_f =c_3 \left(s_{\beta 1}+c_{\beta 1}\kappa_f\right)- s_2 s_3\left(c_{\beta 1} - s_{\beta 1}\kappa_f\right), \nonumber\\
&&y^{h_3}_V = -s_3 s_{\beta 1} - s_2 c_3 c_{\beta 1},~y^{h_3}_f = -s_3 \left(s_{\beta 1}+ c_{\beta 1}\kappa_f\right)-s_2 c_3\left(c_{\beta 1} - s_{\beta 1}\kappa_f\right), \nonumber\\
&&y^{A}_V=0,~y^{A}_{f}=-i\kappa_f~{\rm (for}~u),~y_{f}^{A}=i \kappa_f~{\rm (for}~d,~\ell).
\eea 
Here $\kappa_f=1/\tan\beta$ and $V$ stands for $Z$ or $W$. The shorthand notations are defined as $c_{\beta 1}\equiv\cos(\beta-\alpha_1)$ and $s_{\beta 1}\equiv\sin(\beta-\alpha_1)$.

The charged Higgs Yukawa couplings are given by 
\begin{align} \label{yukawa-charge}
 \mathcal{L}_Y & = - \frac{\sqrt{2}}{v}\, H^+\, \Big\{\bar{u}_i \left[\kappa_f\,(V_{CKM})_{ij}~ m_{dj} P_R
 - \kappa_f\,m_{ui}~ (V_{CKM})_{ij} ~P_L\right] d_j + \kappa_f\,\bar{\nu} m_\ell P_R \ell
 \Big\}+h.c.,
 \end{align}
where $i,j=1,2,3$ are the index of generation.

\section{Relevant theoretical and experimental constraints}
In our discussions, we take into account the following theoretical and experimental constraints:

{\bf (1) Vacuum stability.} The following conditions are required by the vacuum stability \cite{Muhlleitner:2016mzt},
\beq
\Omega_1 ~~{\rm or}~~ \Omega_2
\eeq
with
\begin{align}
\Omega_1 &= \Bigg\{ \lambda_1, \lambda_2, \lambda_6 > 0; \sqrt{\lambda_1 \lambda_6} +
\lambda_7 > 0; \sqrt{\lambda_2 \lambda_6} + \lambda_8 > 0; \nonumber
\\
&\quad \sqrt{\lambda_1 \lambda_2} + \lambda_3 + D > 0; \lambda_7 +
\sqrt{\frac{\lambda_1}{\lambda_2}} \lambda_8 \ge 0 \Bigg\}, \nonumber\\
\Omega_2 &= \Bigg\{ \lambda_1, \lambda_2, \lambda_6 > 0; \sqrt{\lambda_2
  \lambda_6} \ge
\lambda_8 > -\sqrt{\lambda_2 \lambda_6}; \sqrt{\lambda_1 \lambda_6} > -
\lambda_7 \ge \sqrt{\frac{\lambda_1}{\lambda_2}} \lambda_8; \nonumber
\\
&\quad \sqrt{(\lambda_7^2 - \lambda_1 \lambda_6)(\lambda_8^2 -\lambda_2
  \lambda_6)} > \lambda_7 \lambda_8 - (D+\lambda_3) \lambda_6 \Bigg\}
\;.
\label{eq:omega2}
 \end{align}
Here $D = \text{min}(\lambda_4-|\lambda_5|\,,\;0)$.

{\bf (2) Tree-level perturbative unitarity.} The eigenvalues of the $2\to 2$ scalar-scalar scattering matrix are below $8\pi$, which are \cite{Muhlleitner:2016mzt}
\begin{align}
|\lambda_3 - \lambda_4| &< 8 \pi \label{eq:ev1}\nonumber\\
|\lambda_3 + 2 \lambda_4 \pm 3 \lambda_5| &< 8 \pi \nonumber\\
\left| \frac{1}{2} \left( \lambda_1 + \lambda_2 + \sqrt{(\lambda_1 -
     \lambda_2)^2 + 4 \lambda_4^2}\right) \right| &< 8\pi \nonumber\\
\left| \frac{1}{2} \left( \lambda_1 + \lambda_2 + \sqrt{(\lambda_1 -
     \lambda_2)^2 + 4 \lambda_5^2}\right) \right| &< 8\pi \nonumber\\
|\lambda_7|\,,\;|\lambda_8|&<8\pi,  \nonumber\\
\frac{1}{2}|a_{1,2,3}| &< 8\pi \;, 
 \end{align}

where $a_{1,2,3}$ are the real roots of the cubic equation,
\begin{align}
&4\left(-27 \lambda_1 \lambda_2 \lambda_6 + 12 \lambda_3^2 \lambda_6 + 12
\lambda_3 \lambda_4 \lambda_6 + 3 \lambda_4^2 \lambda_6 + 6 \lambda_2
\lambda_7^2 - 8 \lambda_3 \lambda_7 \lambda_8 - 4 \lambda_4 \lambda_7
\lambda_8 + 6 \lambda_1 \lambda_8^2 \right) \nonumber\\
&+ x \left(36 \lambda_1 \lambda_2 -
16\lambda_3^2 - 16\lambda_3 \lambda_4 - 4 \lambda_4^2 + 18 \lambda_1
\lambda_6 + 18 \lambda_2 \lambda_6 - 4\lambda_7^2 - 4\lambda_8^2\right)
\nonumber\\
& +x^2
\left(-6 (\lambda_1 + \lambda_2) -3 \lambda_6\right) + x^3 \;. \label{eq:polynomial}
 \end{align}

{\bf (3) The 125 GeV Higgs signal data.}
We identify $h_2$ as the observed 125 GeV Higgs boson and use $\textsf{HiggsTools}$ \cite{Bahl:2022igd} to compute the total 
$\chi^2_{125}$ based on the latest LHC measurements of signal strength of the 125 GeV Higgs boson. 
$\textsf{HiggsTools}$ integrates both $\textsf{HiggsSignals}$ \cite{Bahl:2022igd,Bechtle:2013xfa} and $\textsf{HiggsBounds}$ \cite{Bechtle:2020pkv,Bechtle:2008jh}.
In particular, we focus on the parameter points that satisfy 
$\chi^2_{125}-\chi^2_{SM}<$ 6.18, where $\chi^2_{SM}$ stands for the SM prediction value.
These points are preferred over the SM at the $2\sigma$ confidence level, assuming a two-parameter fit.

{\bf (4) Searches for additional Higgs boson at the collider and flavor observables.}
We utilize $\textsf{Higgstools}$ to retain only those parameter points that are consistent with the 95\% confidence level exclusion limits
 from collider searches for additional Higgs bosons. We employ $\textsf{SuperIso-v4.1}$ \cite{Mahmoudi:2008tp} to assess the bound of 
$B\to X_s\gamma$ decay.

{\bf (5) The oblique parameters.}
The oblique parameters provide important constraints on the Higgs mass spectrum within the model. 
To ensure consistency with experimental data, parameter points are required to lie within the 
$2\sigma$ confidence region for both the $S$ and $T$ parameters. 
This corresponds to $\chi^2<6.18$ for two degrees of freedom. 
The corresponding fit results can be found in Ref. \cite{pdg2020},
\beq
S=0.00\pm 0.07,~~  T=0.05\pm 0.06, 
\eeq
with a correlation coefficient $\rho_{ST}$ = 0.92.

In the model, the approximate calculations of the $S$ and $T$ parameters are \cite{stu1,stu2}
\bea\label{stu-eq}
S&=&\frac{1}{\pi m_Z^2}\left[\sum_{i=1,2,3}\left( F_S(m_Z^2,m_{h_i}^2,m_A^2) (-s_{\beta} R_{i1} + c_{\beta} R_{i2})^2 \right. \right.\nonumber\\
&&\left.   +F_S(m_Z^2,m_Z^2,m_{h_i}^2) (c_\beta R_{i1} +s_\beta R_{i2})^2 
-m_Z^2 F_{S0}(m_Z^2,m_Z^2,m_{h_i}^2) (c_\beta R_{i1} +s_\beta R_{i2})^2  \right)\nonumber \\
&&\left.  - F_S(m_{Z}^2,m_{H^{\pm}}^2,m_{H^{\pm}}^2) -F_S(m_Z^2,m_Z^2,m_{ref}^2)  + m_Z^2 F_{S0}(m_Z^2,m_Z^2,m_{ref}^2)\right], \nonumber\\
T&=&\frac{1}{16\pi m_W^2 s_W^2} \left[\sum_{i=1,2,3}\left( -F_T(m_{h_i}^2,m_{A}^2) (-s_{\beta} R_{i1} + c_{\beta} R_{i2})^2 \right. \right.\nonumber\\
&&\left. + F_T(m_{H^{\pm}}^2,m_{h_{i}}^2) (c_\beta R_{i2} -s_\beta R_{i1})^2  \right. \nonumber \\
&&\left.\left.+ 3 F_T(m_{Z}^2,m_{h_{i}}^2) (c_\beta R_{i1} +s_\beta R_{i2})^2 -3 F_T(m_{W}^2,m_{h_{i}}^2) (c_\beta R_{i1} +s_\beta R_{i2})^2\right. \right)\nonumber \\
&&\left.     
 + F_T(m_{H^{\pm}}^2,m_A^2) -3 F_T(m_{Z}^2,m_{ref}^2) + 3F_T(m_{W}^2,m_{ref}^2)\right], 
\eea
with
\bea\label{stu-ft}
F_T(a,b)&=&\frac{1}{2}(a+b)-\frac{ab}{a-b}\log(\frac{a}{b}),~~F_S(a,b,c)=B_{22}(a,b,c)-B_{22}(0,b,c),\nonumber\\
F_{S0}(a,b,c)&=&B_{0}(a,b,c)-B_{0}(0,b,c)
\eea
where
\bea
&&B_{22}(a,b,c)=\frac{1}{4}\left[b+c-\frac{1}{3}a\right] - \frac{1}{2}\int^1_0 dx~X\log(X-i\epsilon),\nonumber\\
&&B_{0}(a,b,c) = -\int^1_0 dx ~ \log{(X-i\epsilon)}\, ,\nonumber\\
&&
X=bx+c(1-x)-ax(1-x).
\eea

\section{The diphoton and $b\bar{b}$ excesses at 95.4 GeV} 
The combined analysis of ATLAS and CMS diphoton data at 95.4 GeV reveals a $3.1\sigma$ local excess \cite{Biekotter:2023oen}.
\beq
\mu_{\gamma\gamma}^{exp}=\mu_{\gamma\gamma}^{ATLAS+CMS}=0.24^{+0.09}_{-0.08}.
\eeq
Intriguingly, a persistent local excess with a significance of $2.3\sigma$ was also observed in the $e^+ e^- \to Z(\phi \to b\bar{b})$ channel at LEP in the same mass region \cite{LEPWorkingGroupforHiggsbosonsearches:2003ing},
\beq
\mu_{b\bar{b}}^{exp}=0.117\pm 0.057.
\eeq

The observed excesses in the diphoton and $b\bar{b}$ channels at 95.4 GeV are attributed to resonant production of the lightest CP-even Higgs boson, $h_1$. 
Under the narrow width approximation, the corresponding signal strengths can be formulated as follows:
\begin{align}
	\mu_{b\bar{b}}&=\frac{\sigma_\text{N2HDM}(e^+ e^-\rightarrow Z h_1)}{\sigma_\text{SM}(e^+ e^-\rightarrow Z h_{95.4}^\text{SM})}\times\frac{\text{BR}_\text{N2HDM}(h_1\rightarrow b\bar{b})}{\text{BR}_\text{SM}(h_{95.4}^\text{SM}\rightarrow b\bar{b})}=|y^{h_1}_V|^2\frac{\text{BR}_\text{N2HDM}(h_1\rightarrow b\bar{b})}{\text{BR}_\text{SM}(h_{95.4}^\text{SM}\rightarrow b\bar{b})},
	\label{eq:mulep}
	\end{align}
	\begin{align}
	\mu_{\gamma\gamma}&=\frac{\sigma_\text{N2HDM}(g g\rightarrow h_1)}{\sigma_\text{SM}(g g\rightarrow h_{95.4}^\text{SM})}\times \frac{\text{BR}_\text{N2HDM}(h_1\rightarrow \gamma\gamma)}{\text{BR}_\text{SM}(h_{95.4}^\text{SM}\rightarrow \gamma\gamma)}\simeq |y^{h_1}_f|^2\frac{\text{BR}_\text{N2HDM}(h_1\rightarrow \gamma\gamma)}{\text{BR}_\text{SM}(h_{95.4}^\text{SM}\rightarrow \gamma\gamma)}.
	\label{eq:mucms}
\end{align}
To assess the capability to simultaneously account for the observed excesses in the $\gamma\gamma$ and $b\bar{b}$ channels, 
we conduct a $\chi^2_{95}$ analysis, 
\beq
\chi^2_{95}=\frac{(\mu_{\gamma\gamma}-0.24)^2}{0.085^2}+\frac{(\mu_{b\bar{b}}-0.117)^2}{0.057^2},
\eeq
and explain the diphoton and $b\bar{b}$ excesses within the $2\sigma$ ranges, namely $\chi^2_{95}< 6.18$.

In the model, the decay $h_1 \to \gamma\gamma$ is induced at the one-loop level, whose width is calculated as
\begin{equation}
    \begin{array}{lllll}
    \Gamma(h_1 \to \gamma\gamma) & = &  \displaystyle
            \frac{\alpha^2 m_{h_1}^3}{256 \pi^3 v^2}
            \left| y_V^{h_1}  F_1(\tau_{W^\pm}) + y_{H^\pm} F_0(\tau_{H^\pm})+\sum_f y_f^{h_1} N_{cf} Q_f^2 F_{1/2}(\tau_f) \right|^2,
    \end{array}
\label{widthrr}
\end{equation}
where $\tau_i=\frac{4m^2_{i}}{m^2_{h_1}}$, and $N_{cf}$ and $Q_f$ are electric charge and the number of color degrees of freedom of the
fermion in the loop. The factor $y_{H^\pm}$ is 
\beq
y_{H^\pm}=\frac{v}{2m_{h^\pm}^2}g_{h_1H^+H^-}
\eeq
with
\beq
g_{h_1H^+H^-}=\frac{1}{v} \left( - \frac{m_{12}^2}{s_\beta c_\beta} \left[
   \frac{R_{11}}{c_\beta} + \frac{R_{12}}{s_\beta} \right] +
 m_{h_1}^2 \left[ \frac{R_{11} s_\beta^2}{c_\beta} + \frac{R_{12}
     c_\beta^2}{s_\beta} \right] + 2m_{H^\pm}^2 [R_{11} c_\beta + R_{12}
 s_\beta] \right)
\eeq
The functions $F_{1,0,1/2}$ are defined as \cite{Djouadi:2005gj}
\beq
    F_1(\tau) = 2 + 3 \tau + 3\tau (2-\tau) f(\tau),\quad
    F_{1/2}(\tau) = -2\tau [1 + (1-\tau)f(\tau)],\quad
    F_0(\tau) = \tau [1 - \tau f(\tau)],
\eeq
with
\begin{equation}
    f(\tau) = \left\{ \begin{array}{lr}
        [\sin^{-1}(1/\sqrt{\tau})]^2, & \tau \geq 1 \\
        -\frac{1}{4} [\ln(\eta_+/\eta_-) - i \pi]^2, & \, \tau < 1
        \end{array}  \right.\label{hggf12}
\end{equation}
where $\eta_{\pm}=1\pm\sqrt{1-\tau}$. 

In our calculations, we take two different scenarios:

{\bf Scenario A:} By choosing a small value of $c_2$, 
the 95.4 GeV Higgs boson ($h_1$) is predominantly singlet-like, originating mainly from the singlet field $\Phi_S$.
Under the approximation of $s_2\simeq sgn(s_2)(1-\frac{c_2^2}{2})$, the couplings $h_2$ and $h_3$ normalized to the SM are given by
\bea\label{hffcoupling-a}
&&y^{h_2}_V \simeq |s_2| s_{\beta 13}+\frac{c_2^2}{2}c_3 s_{\beta 1},~y^{h_2}_f \simeq |s_2| \left(s_{\beta 13}+c_{\beta 13}\kappa_f\right)+\frac{c_2^2}{2}c_3\left(s_{\beta 1} + c_{\beta 1}\kappa_f\right), \nonumber\\
&&y^{h_3}_V \simeq |s_2| c_{\beta 13}-\frac{c_2^2}{2}c_3 s_{\beta 1},~y^{h_3}_f \simeq |s_2| \left(c_{\beta 13}-s_{\beta 13}\kappa_f\right)-\frac{c_2^2}{2}c_3\left(s_{\beta 1} + c_{\beta 1}\kappa_f\right), \eea 
where $c_{\beta 13}\equiv\cos(\beta-\alpha_1-sgn(s_2)\alpha_3)$ and $s_{\beta 13}\equiv\sin(\beta-\alpha_1-sgn(s_2)\alpha_3)$.
The mixing parameters are chosen as follows: 
\beq\label{parameter}
 0 \leq c_2 \leq 0.4,~~ s_{\beta 13} = 1.0,~~
  0 \leq c_{\beta 1} \leq 1.0,~~ 1 \leq \tan\beta \leq 15.
\eeq
When $s_{\beta 13} = 1.0$ and $c_2$ approaches zero, the couplings of $h_2$ to the SM particles converge to 
those of the SM Higgs boson, which is favored by the observed signal data for the 125 GeV Higgs.

{\bf Scenario B:} The mixing parameters are scanned over the following ranges:
\beq\label{parameter}
  0 \leq c_{\beta 1} \leq  0.4,~~-0.4 \leq s_{3} \leq 0.4,~~0 \leq c_2 \leq 1,~~ 1 \leq \tan\beta \leq 15.
\eeq
In the limits of $c_{\beta 1} \to 0$ and $|s_3| \to 0$, the couplings of $h_2$ to the SM particles approach those of the SM,
 which is favored by the signal measurements of the 125 GeV Higgs.
In addition, when $c_2 \to 1$, the 95.4 GeV Higgs boson ($h_1$) predominantly originates from the mixing of the two CP-even components of the Higgs doublets.

\begin{figure}[tb]
\centering
\includegraphics[width=16.5cm]{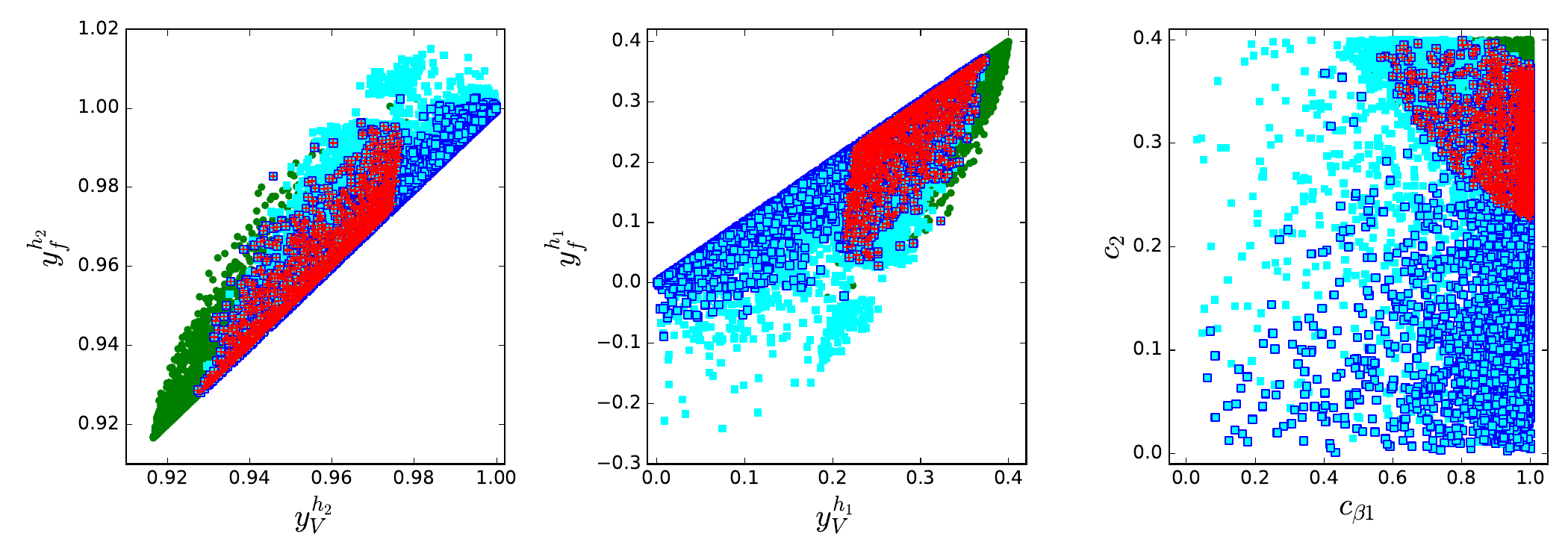}
\vspace{-1.3cm}\caption{In scenario A, the parameter space is progressively constrained by sequentially imposing the following: 
theoretical requirements and the oblique parameters; the signal data of the 125 GeV Higgs boson; 
searches for additional Higgs bosons at the collider and flavour observables; and finally the condition $\chi_{95}^2<6.18$. 
The surviving parameter points at each stage are represented by green bullets, cyan squares, blue-edged squares, and red pluses, respectively.}
\label{figalhc}
\end{figure}

\begin{figure}[tb]
\centering
\includegraphics[width=17cm]{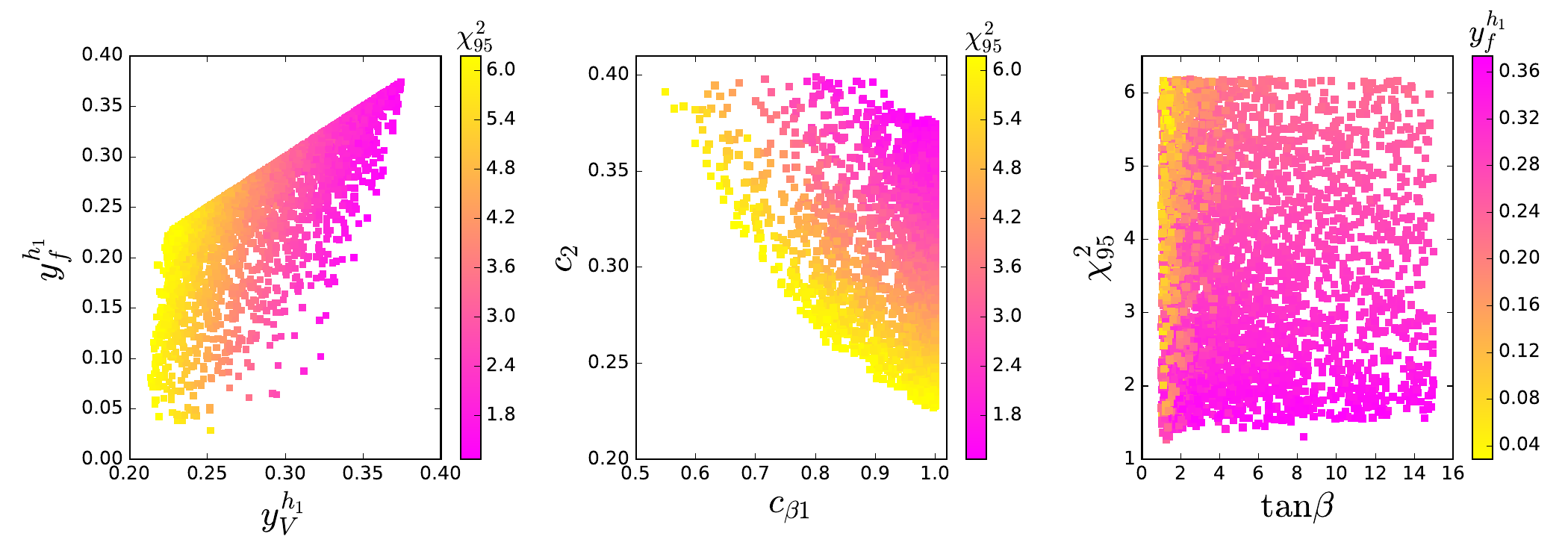}
\vspace{-1.3cm}\caption{In scenario A, the surviving parameter points satisfy $\chi_{95}^2<6.18$ while remaining consistent with
 the theoretical constraints, the oblique parameters, the Higgs signal data at 125 GeV, 
searches for additional Higgs bosons at the collider, and flavor observables.}
\label{figachi}
\end{figure}

\begin{figure}[tb]
\centering
\includegraphics[width=16.5cm]{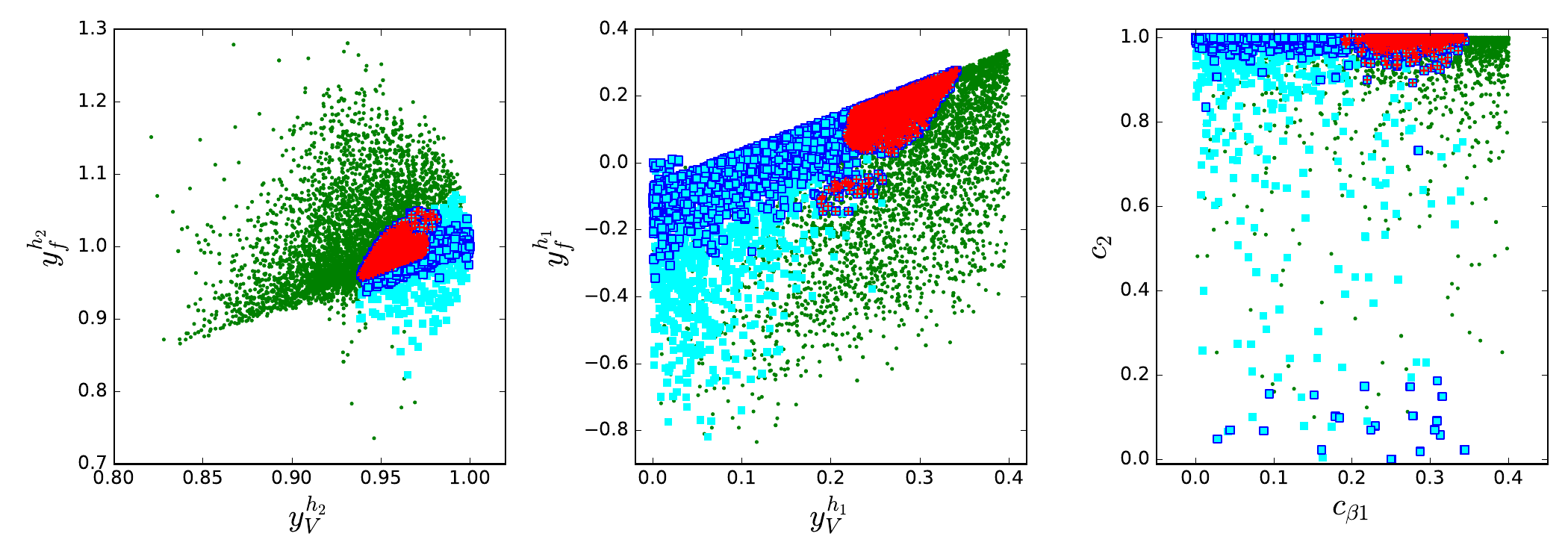}
\vspace{-1.3cm}\caption{Same as the Fig. \ref{figalhc}, but for scenario B.}
\label{figblhc}
\end{figure}
 
\begin{figure}[tb]
\centering
\includegraphics[width=14.cm]{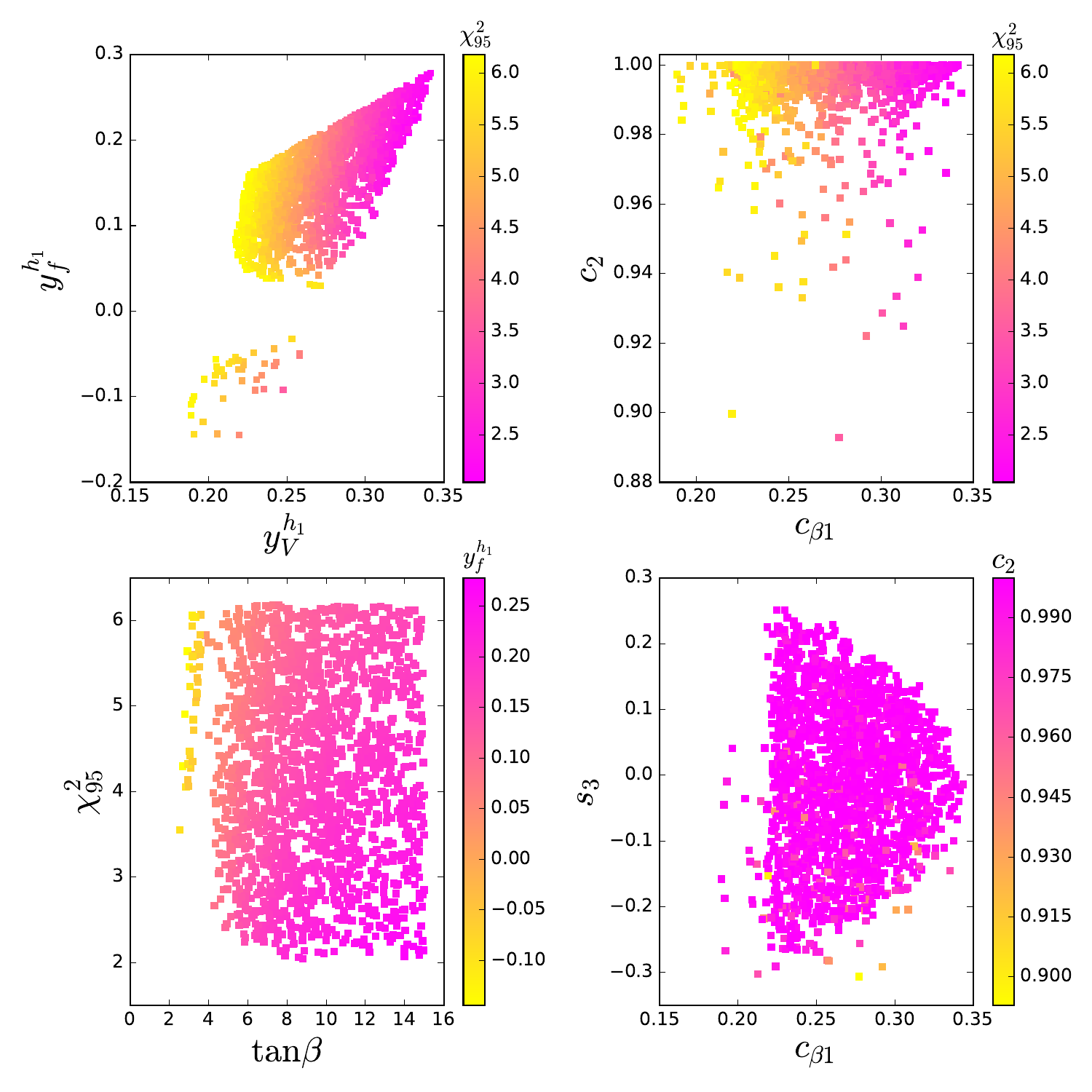}
\vspace{-0.8cm}\caption{Same as the Fig. \ref{figachi}, but for scenario B.}
\label{figbchi}
\end{figure}

After imposing the relevant theoretical constraints, the oblique parameters, the Higgs signal data at 125 GeV, 
searches for additional Higgs bosons at the collider, and flavor observables, 
we identify the parameter space in scenario A that accounts for the observed excesses in the $b\bar{b}$ and diphoton channels at 95.4 GeV in Fig. \ref{figalhc}.
As indicated by Eq.~(\ref{hffcoupling-a}), within the selected parameter space of scenario A, the coupling of $h_2$ to vector bosons, $y_V^{h_2}$, 
tends to be smaller than its coupling to fermions, $y_f^{h_2}$, as illustrated in the left panel. 
The Higgs signal data at 125 GeV impose a lower bound on $y_V^{h_2}$, requiring $y_V^{h_2} > 0.93$. 
In the parameter space consistent with the 125 GeV Higgs measurements, collider limits on additional Higgs bosons, and constraints from flavor physics, 
the coupling $y_V^{h_1}$ is restricted to values below 0.38.
Furthermore, the condition $\chi^2_{95} < 6.18$
is satisfied in the region of $0.2 \le y_V^{h_1} \le 0.38$, with $y_f^{h_1}$ bounded from below.
In addition, the Higgs signal data at 125 GeV exclude a small corner of the parameter space characterized by $c_{\beta 1}\to 1$ and $c_2\to 0.4$.
In such region both $y_V^{h_2}$ and $y_f^{h_2}$ can exhibit significant deviations from unity.

As shown in the left panel of Fig. \ref{figachi}, the value of $\chi_{95}^2$ is sensitive to $y^{h_1}_V$ and $y^{h_1}_f$,
 and decreases as they increase. The condition $\chi_{95}^2<2.3$ is satisfied in partial region of parameter space, with the minimum value attaining 1.2,
which indicates that the model can explain the diphoton and $b\bar{b}$ excesses within the $1\sigma$ ranges. 
The signal strength $\mu_{b\bar{b}}$ is proportional to $|y^{h_1}_V|^2$.
Meanwhile, the $W$-loop can play an important contribution to the width of $h_1\to \gamma\gamma$, and
therefore the signal strength $\mu_{\gamma\gamma}$ is also affected by $y^{h_1}_V$.
The interpretation of the $b\bar{b}$ excess imposes a lower bound on $|y_f^{h_1}|$.
Because of $y_V^{h_1}=c_2c_{\beta 1}$, the value of $\chi_{95}^2$ tends to decrease as $c_2$ and $c_{\beta 1}$ increase,
as shown in the middle panel of Fig. \ref{figachi}. 
When $c_{\beta 1}$ or $\tan\beta$ is large, the $y^{h_1}_f$ will has a mild dependence on $\tan\beta$ (See Eq. (\ref{hffcoupling})).
From the right panel of Fig. \ref{figachi}, it is observed that the value of $\chi_{95}^2$ shows limited sensitivity to $\tan\beta$.

Similar to Fig. \ref{figalhc}, we display the parameter points for scenario B that accounts for
 the observed excesses in the $b\bar{b}$ and diphoton channels at 95.4 GeV in Fig. \ref{figblhc}. 
Unlike scenario A, the allowed region in scenario B includes not only the positive $y_f^{h_1}$ region 
but also a narrow band where $y_f^{h_1}$ takes negative values. The signal data of the 125 GeV Higgs boson
requires a small $c_{\beta 1}$,  allowing for a negative $y_f^{h_1}$, 
see Eq. (\ref{hffcoupling}). The condition $\chi_{95}^2 < 6.18$ is satisfied in the region of $0.18<c_{\beta 1}< 0.35$ and $0.88<c_2<1$.
For such large $c_2$, the 95.4 GeV Higgs boson mainly originates from the mixing of the two CP-even components of the Higgs doublets.

As in scenario A, the value of $\chi_{95}^2$ in scenario B decreases  as $y^{h_1}_V$ and $\mid y^{h_1}_f\mid$ ($c_2$ and $c_{\beta 1}$) increase, 
as illustrated in the upper-left panel and upper-right panel of Fig. \ref{figbchi}. However, different from scenario A, the minimal value
of $\chi_{95}^2$ in scenario B only reaches to 2, suggesting that the model can only marginally explain 
the diphoton and $b\bar{b}$ excesses at the $1\sigma$ level.
Furthermore, the condition $\chi_{95}^2 < 6.18$ disfavors a narrow region around $\tan\beta\sim 4$, 
where the value of $\mid y_f^{h_1}\mid$ is significantly suppressed.  
The parameter $s_3$ is favored to vary from -0.3 to 0.25, and is not sensitive to $c_2$. 
Its absolute value decreases with increasing $c_{\beta 1}$ in order to maintain a sufficiently large $y_V^{h_2}$ to accommodate the signal data of the
125 GeV Higgs boson.

\section{Domain walls and gravitational waves}
The classical scalar fields are parameterized as
\bea
&&\Phi_1=\left(\begin{array}{c} 0 \\
\frac{\phi_1(z)}{\sqrt{2}}\,
\end{array}\right)\,, 
\Phi_2=\left(\begin{array}{c} 0 \\
\frac{\phi_2(z)}{\sqrt{2}}\,
\end{array}\right),
\Phi_S=\phi_s(z),
\eea
where $z$ is a coordinate perpendicular to domain walls \cite{Hattori:2015xla}. The expression for the energy density of the domain walls is given by
\beq
\varepsilon_{\rm wall}=\frac{1}{2}(\partial_z\phi_1)^2 + \frac{1}{2}(\partial_z\phi_2)^2 + \frac{1}{2}(\partial_z\phi_S)^2 + V(\phi_1,\phi_2,\phi_S)
-V(v_1,v_2,v_s),
\eeq
with the scalar potential
\begin{equation}
  \begin{aligned}
  V(\phi_1,\phi_2,\phi_S)&=\frac{m_{11}^2}{2} \phi_1^2 +\frac{\lambda_1}{8} \phi_1^4+ \frac{m_{22}^2}{2} \phi_2^2+\frac{\lambda_2}{8} \phi_2^4-m_{12}^2 \phi_1 \phi_2+\frac{\lambda_{345}}{4} \phi_1^2  \phi_2^2  \\ 
  &\hphantom{=}+\frac{m_S^2}{2} \phi_S^2 +\frac{\lambda_6}{8} \phi_S^4+ \frac{\lambda_7}{4}  \phi_1^2\phi_S^2 +\frac{\lambda_8}{4} \phi_2^2 \phi_S^2-a_1 v_s^2 \phi_S+\frac{a_1}{3} \phi_S^3.
  \end{aligned}
\end{equation}

The equations of motion for domain walls are
\begin{equation}
\label{eq:bubble_equation}
\frac{ \mathrm{d}^2 \phi_i }{ \mathrm{d} z^2 }  = \frac{ \partial V(\phi_1,\phi_2,\phi_S) }{ \partial \phi_i }, \quad ( \phi_i=\phi_1,\phi_2,\phi_S), 
\end{equation}
with the boundary conditions
\begin{align}
\lim_{z\to\pm\infty}\phi_1(z)&=v_1,\quad \lim_{z\to\pm\infty}\phi_2(z)=v_2, \quad \lim_{z\to\pm\infty}\phi_S(z)=\pm v_s.\label{eom_dw_bc}
\end{align}
We modify $\textsf{CosmoTransitions}$ \cite{cosmopt} to approximately solve Eqs.~(\ref{eq:bubble_equation}) and (\ref{eom_dw_bc}), and 
relevant computational details are provided in \cite{cosmopt,Chen:2020wvu}.
The domain wall tension $\sigma_{\rm wall}$ is computed through an integration of the energy density 
$\varepsilon_{\rm wall}$ over the $z$ axis,
\beq
\sigma_{\rm wall}=\int dz ~\varepsilon_{\rm wall}.
\eeq

Domain walls must collapse before they can overclose
the Universe. The volume pressure induced by  energy bias ($V_{\rm bias}$) exerts a force on the domain walls, and 
drives the contraction of regions occupied by the false vacuum.  The collapse of domain wall occurs when the volume pressure force
becomes comparable to the tension force. The characteristic temperature at which the domain walls annihilate can be estimated by \cite{Hiramatsu:2013qaa,Saikawa:2017hiv}
\beq\label{eq:Tann}
T_{\rm ann} =3.41\times 10^{-2}\,{\rm GeV}\, C_{\rm ann}^{-1/2} A^{-1/2} \Big(  \frac{ g_{*} (T_{\rm ann})  }{10}  \Big)^{-1/4}  \hat \sigma_{\rm wall}^{-1/2} \hat V_{\rm bias}^{1/2} \,,
\eeq
with
\beq
\hat \sigma_{\rm wall} \equiv \frac{\sigma_{\rm wall}}{\rm TeV^3},~~ \hat V_{\rm bias} \equiv \frac{V_{\rm bias}}{\rm MeV^4}.
\eeq
In this analysis, we adopt a constant $C_{ann}=5$ \cite{Saikawa:2017hiv} and set the area parameter $A\simeq 0.8$ for the $Z_2$ symmetric model \cite{Hiramatsu:2013qaa}.
The energetic particles released from the collapse of domain walls have the potential to disrupt the synthesis of light nuclei formed during Big Bang Nucleosynthesis (BBN) \cite{Kawasaki:2004qu}.
To preserve the successful predictions of BBN, it is therefore imperative that the domain walls annihilate before the
BBN epoch, $T_{\rm ann} > T_{\rm BBN} = 8.6 \times 10^{-3}$ GeV. In addition, 
in order to ensure that domain walls annihilate before their energy density dominates the Universe, a lower bound must be imposed on $V_{\rm bias}$ \cite{Saikawa:2017hiv}
\begin{equation}\label{eq:45}
V_{\rm{bias}}^{1/4}>2.18\times 10^{-5} \hspace{1mm} C_{\rm{ann}}^{\frac{1}{4}} \hspace{1mm} A^{\frac{1}{2}} \hat \sigma_{\rm wall}^{\frac{1}{2}}\,\,.
\end{equation}



Following the annihilation of domain walls, a SGWB is produced, and the peak frequency is given by \cite{Hiramatsu:2013qaa}
\bea\label{eq:fpeak}
f_{\rm peak}& \simeq & 1.1\times 10^{-7}\,{\rm Hz}\, \Big(  \frac{ g_*( T_{\rm ann} ) }{10}  \Big)^{1/2}  \Big(  \frac{ g_{*s}( T_{\rm ann} ) }{10}  \Big)^{-1/3} \Big(  \frac{ T_{\rm ann} }{1\,\rm GeV }  \Big)\nonumber\\
&\simeq&  3.75\times 10^{-9}\,{\rm Hz}\,  \Big(  \frac{ g_*( T_{\rm ann} ) }{10}  \Big)^{1/4} \Big(  \frac{ g_{*s}( T_{\rm ann} ) }{10}  
\Big)^{-1/3} C_{\rm ann}^{-1/2} A^{-1/2} \hat\sigma^{-1/2}_{\rm wall}  
 \hat V_{\rm bias}^{1/2}    \,.
\eea
Here, $g_*(T_{\rm ann})$ and $g_{*s}(T_{\rm ann})$ count the relativistic degrees of freedom contributing to the energy density and the entropy density. 
Within the temperature range $1\,\rm MeV\lesssim T_{\rm ann} \lesssim 100\,\rm MeV$, both quantities take the value 10.75 \cite{Husdal:2016haj}.

The peak amplitude of the GW spectrum at the present time is \cite{Hiramatsu:2013qaa,Saikawa:2017hiv}
\beq\label{gw-peak}
\Omega_{\rm GW}^{\rm peak} h^2 =  5.3\times 10^{-20}\, \tilde \epsilon_{\rm GW} A^4 C_{\rm ann}^2 \Big( \frac{ g_{*s}(T_{\rm ann})  }{10}   \Big)^{-4/3}  
 \Big(  \frac{ g_{*} (T_{\rm ann})  }{10}  \Big) \hat \sigma^4_{\rm wall} \hat V^{-2}_{\rm bias}  \,.
\eeq
with $\tilde \epsilon_{\rm GW}\simeq 0.7\pm 0.4$ \cite{Hiramatsu:2013qaa}.
Given an arbitrary frequency $f$, we use \cite{Hiramatsu:2013qaa,Hiramatsu:2010yz}
\begin{align}
\Omega_{\text{GW}}h^2(f<f_{\text{peak}}) &= \Omega_{\rm GW}^{\rm peak} h^2 \left(\frac{f}{f_{\text{peak}}}\right)^3, \\
\Omega_{\text{GW}}h^2(f>f_{\text{peak}}) &= \Omega_{\rm GW}^{\rm peak} h^2 \left(\frac{f_{\text{peak}}}{f}\right).
\end{align}

\begin{figure}[tb]
\centering
\includegraphics[width=13.cm]{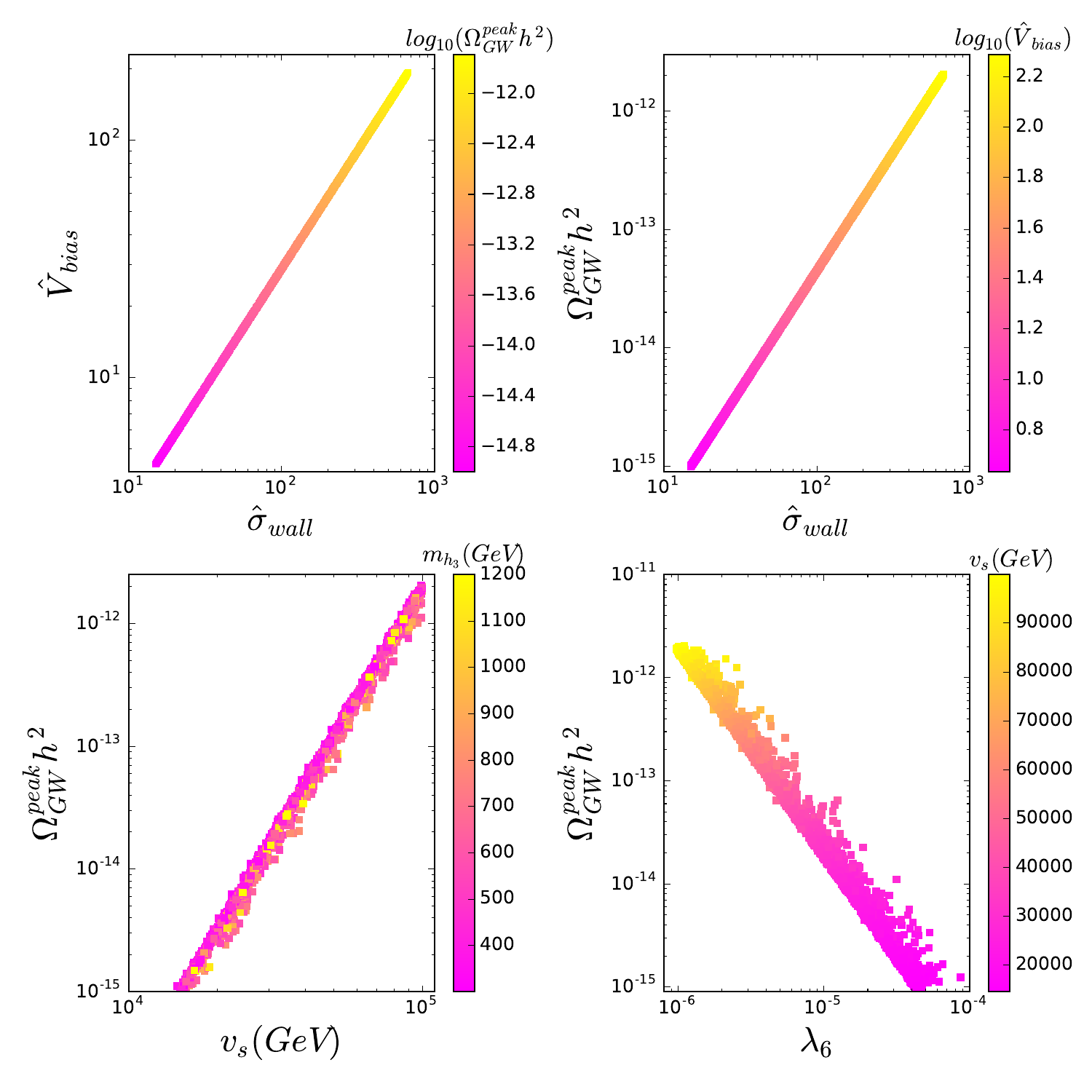}
\vspace{-0.7cm}\caption{In scenario A, the parameter points satisfy $\Omega_{\rm GW}^{\rm peak} h^2>10^{-15}$ and $\chi_{95}^2<6.18$ 
while remaining consistent with relevant theoretical and experimental constraints mentioned above.}
\label{figagw}
\end{figure}
 
\begin{figure}[tb]
\centering
\includegraphics[width=16.cm]{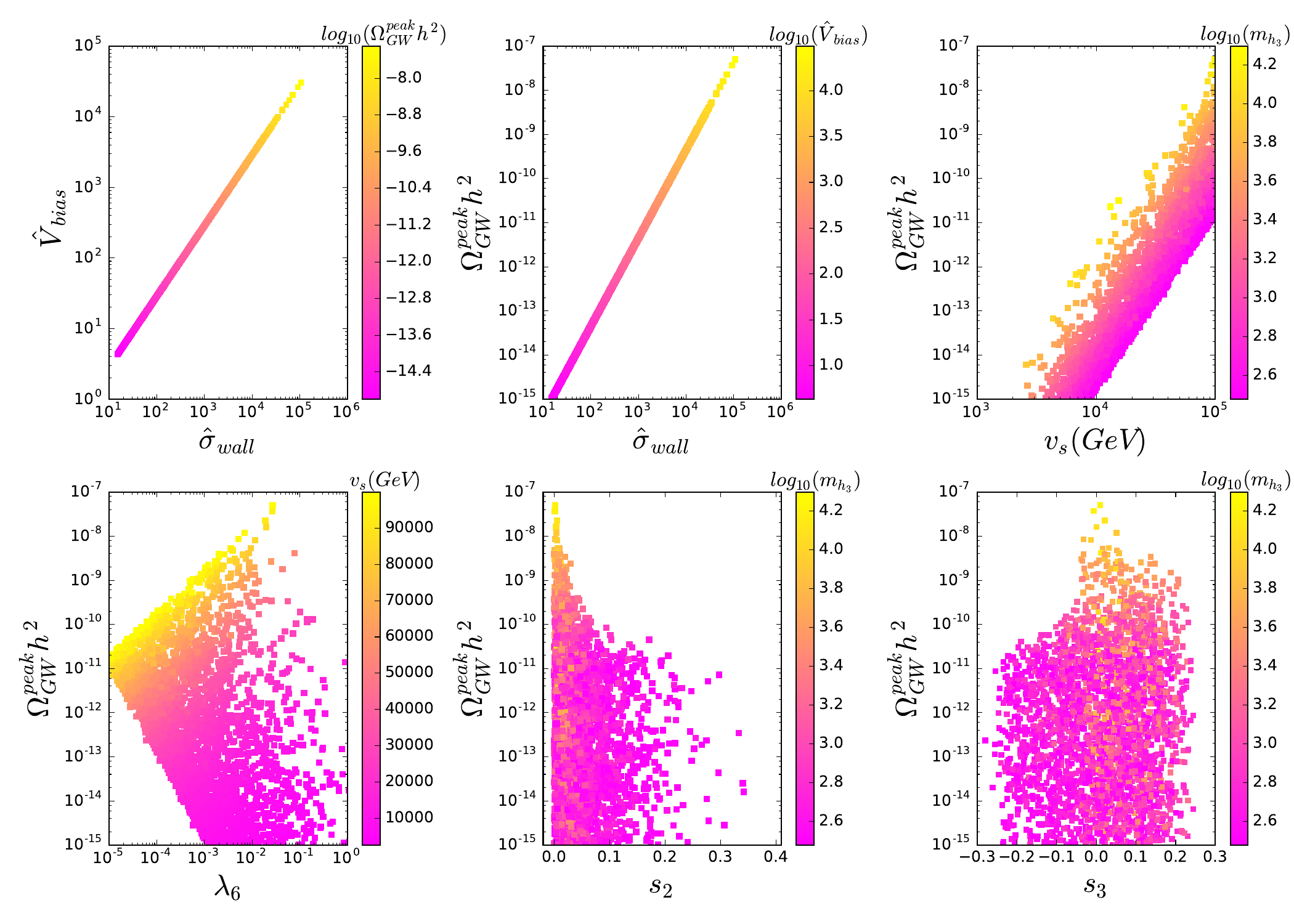}
\vspace{-0.7cm}\caption{Same as the Fig. \ref{figagw}, but for scenario B.}
\label{figbgw}
\end{figure}

\begin{figure}[tb]
\centering
\includegraphics[width=9.cm]{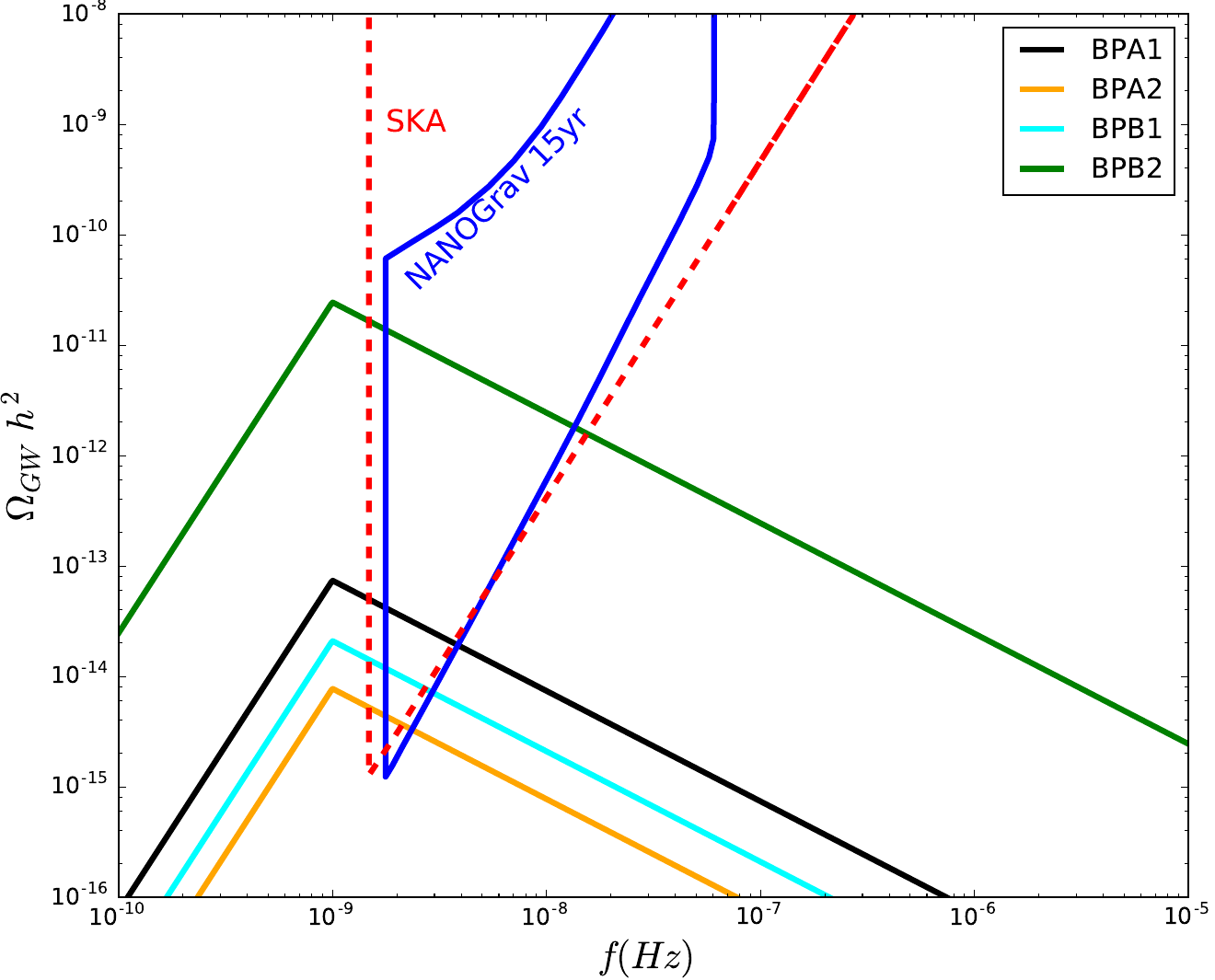}
\vspace{-0.4cm}\caption{The GW spectra $\Omega_{\rm GW}h^2$ as a function of the frequency $f$.
The result of the NANOGrav 15-year dataset \cite{NANOGrav:2023gor,NANOGrav:2023hfp} and the sensitivity
of SKA \cite{Carilli:2004nx} are also displayed.}
\label{figska}
\end{figure}

In our discussions, we choose $f_{\rm peak}$ around $10^{-9}$ Hz to get $\hat V_{\rm bias}$ from the second line of Eq. (\ref{eq:fpeak}) with 
$\hat \sigma_{\rm wall}$ calculated in the parameter points accommodating the diphoton and $b\bar{b}$ excesses at 95.4 GeV. 
From the first line of Eq. (\ref{eq:fpeak}), we obtain $T_{\rm ann}\simeq 9$ MeV for $f_{\rm peak}=10^{-9}$ Hz. 
In Fig. \ref{figagw} and Fig. \ref{figbgw},  we  display the parameter points that satisfy $\Omega_{\rm GW}^{\rm peak} h^2>10^{-15}$ and $\chi_{95}^2 < 6.18$ 
while remaining consistent with relevant theoretical and experimental constraints mentioned above, for scenario A and scenario B, respectively.

For a fixed $f_{\rm peak}$, both $\hat V_{\rm bias}$ and $\Omega_{\rm GW}^{\rm peak}$ increase with $\sigma_{\rm wall}$, 
as indicated by the second line of Eq. (\ref{eq:fpeak}) and Eq. (\ref{gw-peak}), and as illustrated in the upper-left and upper-right panels of Fig. \ref{figagw}.
In scenario A, although the 95.4 GeV Higgs boson ($h_1$) predominantly originates from the real singlet field, it must include non-negligible components 
of the Higgs doublets to acquire the necessary couplings to gauge bosons and fermions, thereby accounting for the observed diphoton and
 and $b\bar{b}$ excesses at 95.4 GeV. Thus,
 both $R_{31}$ and $R_{32}$ can not approach to zero, theoretical constraints on $\lambda_{1,2,3}$ will impose a lower bound on $m_{h_3}$,
$m_{h_3}<1.2$ TeV (see the first three lines of Eq. (\ref{eq:lambdas}) ). As a result, according to the sixth line of Eq. (\ref{eq:lambdas}), $\lambda_6$ 
decreases  with increasing $v_s$, yielding a range of
$10^{-6}<\lambda_6<10^{-4}$ for $14~{\rm TeV}~<v_s<100$ TeV.
 As $v_s$ increases, although $\lambda_6$ decreases, $\sigma_{\rm wall}$ still grows rapidly, resulting in $\Omega_{\rm GW}^{\rm peak}h^2$ reaching 
$2\times 10^{-12}$ for $v_s=100$ TeV and $\lambda_6$ around $10^{-6}$.

In scenario B, the 95.4 GeV Higgs boson predominantly originates from the CP-even components of the Higgs doublets.
Even though the mixing  between the singlet and doublet components vanishes (i.e., $R_{31}\to 0$ and $R_{32}\to 0$), the 95.4 GeV Higgs boson
still can interact with gauge bosons and fermions to explain the diphoton 
 and $b\bar{b}$ excesses at 95.4 GeV. Therefore, theoretical constraints on $\lambda_{1,2,3}$ allow
$m_{h_3}$ to have a large value, reaching up to 20 TeV in the limits $s_2\to 0$ and $s_3\to $0
as illustrated in the lower-middle panel and lower-right panel of Fig. \ref{figbgw}.
As a result, $\lambda_6$ is no longer suppressed by a large $v_s$,
and can reach 0.03 for $v_s=100$ TeV.
As shown in the Fig. \ref{figbgw}, the peak amplitude of the GW spectrum, $\Omega_{\rm GW}^{\rm peak}h^2$, can reach $6\times 10^{-8}$ for
$v_s=100$ TeV, $\lambda_6=0.03$, $s_2\to 0$ and $s_3\to $0.


\begin{table}[t]
\centering
\begin{tabular}{| c | c | c | c | c |c | c | c | c |c | c | c | c | c | c | c |c | c | c | c |c |c |c |c |}
\hline
& $m_{h_3}$(GeV) & $m_{A}=m_{H^\pm}$(GeV) & $m_{12}^2$(GeV$)^2$& $\tan\beta$& $s_{1}$& $s_{2}$& $s_{3}$ \\
\hline
BPA1   & 455.71 &  623.58  & 35771.69 & 1.37 & 0.2875& 0.9174& -0.7973\\
 \hline
BPA2    & 441.82   & 549.95  & 28387.58 & 1.85 & 0.4013& 0.9165& -0.7890 \\
 \hline
BPB1    & 3202.30&  331.47  & 683.50 & 13.09 & 0.2108& 0.0406& 0.1644\\
 \hline
BPB2   & 3349.33& 431.49  & 649.92 & 14.76 & 0.2633& 0.0260& 0.0569 \\
 \hline
\end{tabular}
\centering
\begin{tabular}{| c | c | c| c | c | c |c | c | c | c |c | c | c | c |c | c | c | c | c | c | c |c | c | c | c |c |c |}
\hline
 &$v_s$(GeV)~&$\lambda_6$ &$~\hat \sigma_{wall}$~~&~$\hat V_{bias}$~& $\Omega^{peak}_{GW}h^2$     &  $~\chi^2_{95}$ \\
\hline
BPA1& 44917.62& $1.06\times 10^{-5}$&128.05 &$36.97$& ~$7.39\times 10^{-14}$~      & 1.99\\
 \hline
BPA2& 24174.21&$3.61\times 10^{-5}$ &41.45  & $11.97$&~$7.75\times 10^{-15}$~       & 2.22\\
 \hline
BPB1& 5721.32&$3.04\times 10^{-1}$&68.20 &$19.69$& ~$2.10\times 10^{-14}$~     & 3.78\\
 \hline
BPB2& 32336.01&$1.07\times 10^{-2}$&2328.69 & $672.3$&~$2.45\times 10^{-11}$~   & 2.42\\
 \hline
\end{tabular}
\caption{Input parameters and results for the BPA1, BPA2, BPB1 and BPB2. Here $m_{h_1}=95.4$ GeV and $m_{h_2}=125.5$ GeV are fixed.}
\label{tabbp1}
\end{table}

We pick out four benchmark points: BPA1 and BPA2 for scenario A, and BPB1 and BPB2 for scenario B, and display the key input parameters and results
 in Table \ref{tabbp1}.  
We examine the GW spectra for the four benchmark points, which are shown along with the result of the NANOGrav 15-year dataset and the sensitivity curve
of SKA in Fig. \ref{figska}.
We find that the GW spectra characterized by $f_{\rm peak}=10^{-9}$ Hz and $\Omega_{\rm GW}^{\rm peak}h^2> \order(10^{-14})$
can explain the current PTA GW data and be confirmed by upcoming
PTA such as SKA. 
As discussed above, although the peak amplitude of the GW spectra in scenario B can be significantly larger than
that in scenario A, the former only marginally accounts for
the diphoton and $b\bar{b}$ excesses at the $1\sigma$ level, whereas the latter can fully explain both excesses within the $1\sigma$ range.

\section{Conclusion}

After imposing relevant theoretical and experimental constraints, we studied the diphoton and $b\bar{b}$ excesses at 95.4 GeV, 
along with the nano-Hertz GW signatures originating from domain walls, within the framework of the N2HDM.
When the 95.4 GeV Higgs boson predominantly originates from the singlet field, we found that the model fully accounts for the diphoton and $b\bar{b}$ excesses at 95.4 GeV within the $1\sigma$ range.
Additionally, this scenario can predict a GW spectrum with a peak amplitude of $2\times 10^{-12}$ at a peak frequency of $10^{-9}$ Hz for $v_s=100$ TeV.
When the 95.4 GeV Higgs boson arises mainly from the CP-even components of the Higgs doublets, we found that the scenario only marginally accounts for
the diphoton and $b\bar{b}$ excesses at the $1\sigma$ level, but can yield a significantly enhanced GW amplitude of $6\times 10^{-8}$ at the same frequency for the same value of
$v_s$.
The nano-Hertz GW signals predicted by the model can explain the current PTA GW data and be confirmed by upcoming
PTA such as SKA.
The viable parameter space corresponding to each scenario have been explicitly mapped out.
Our works highlight the complementary roles of collider signals and GW observables in probing the nature of the extended Higgs sector.

\section*{Acknowledgment}
We thank Yang Zhang and Yeling Zhou for helpful discussions.
This work was supported by the National Natural
Science Foundation of China under Grant No. 11975013
and by the Projects No. ZR2024MA001 and
No. ZR2023MA038 supported by Shandong Provincial 
Natural Science Foundation. 


\end{document}